\documentclass[conference]{IEEEtran}
\IEEEoverridecommandlockouts
\usepackage{url}
\usepackage{cite}
\usepackage{amsmath,amssymb,amsfonts}
\usepackage{algorithmic}
\usepackage{graphicx}
\usepackage{textcomp}
\usepackage{xcolor}
\usepackage[colorinlistoftodos]{todonotes}
\def\BibTeX{{\rm B\kern-.05em{\sc i\kern-.025em b}\kern-.08em
    T\kern-.1667em\lower.7ex\hbox{E}\kern-.125emX}}

\usepackage{color, soul}

\def\BibTeX{{\rm B\kern-.05em{\sc i\kern-.025em b}\kern-.08em
    T\kern-.1667em\lower.7ex\hbox{E}\kern-.125emX}}

\usepackage{tikz}
\usepackage{hyperref}
\usepackage{lipsum}

\newcommand\copyrighttext{%
  \footnotesize \textcopyright 2025 IEEE. Personal use of this material is permitted.
  Permission from IEEE must be obtained for all other uses, in any current or future 
  media, including reprinting/republishing this material for advertising or promotional 
  purposes, creating new collective works, for resale or redistribution to servers or 
  lists, or reuse of any copyrighted component of this work in other works. 
  }
\newcommand\copyrightnotice{%
\begin{tikzpicture}[remember picture,overlay]
\node[anchor=south,yshift=10pt] at (current page.south) {\fbox{\parbox{\dimexpr\textwidth-\fboxsep-\fboxrule\relax}{\copyrighttext}}};
\end{tikzpicture}%
}
    
\begin{document}

\title{A Model-Based Approach to Automated Digital Twin Generation in Manufacturing \\
\thanks{This work was partially funded by the National Recovery and Resilience Plan Greece 2.0, funded by the European Union – NextGeneration EU, Greece4.0 project, under agreement no. TAEDR-0535864. The paper reflects the authors’ views, and the Commission is not responsible for any use that may be made of the information it contains.}
}

\author{
    \IEEEauthorblockN{Angelos Alexopoulos\IEEEauthorrefmark{1}\IEEEauthorrefmark{2},
    Agorakis Bompotas\IEEEauthorrefmark{1},
    Nikitas Rigas Kalogeropoulos\IEEEauthorrefmark{1},
    Panagiotis Kechagias\IEEEauthorrefmark{1},\\
    Athanasios P. Kalogeras\IEEEauthorrefmark{1},
    Christos Alexakos\IEEEauthorrefmark{1}}
    \\
    \IEEEauthorblockA{\IEEEauthorrefmark{1}\textit{Industrial Systems Institute} \\
    \textit{ATHENA Research Center}\\
    Patras, Greece \\
    \{aggalexopoulos, abompotas, nkalogeropoulos, kechagias, kalogeras, alexakos\}@athenarc.gr}
    \IEEEauthorblockA{\IEEEauthorrefmark{2}\textit{Physics Department} \\
    \textit{University of Patras}\\
    Patras, Greece }
}

\IEEEpubid{\makebox[\columnwidth]{979-8-3315-0448-9/24/\$31.00~\copyright2024 IEEE \hfill}
\hspace{\columnsep}\makebox[\columnwidth]{ }}

\maketitle
\IEEEpubidadjcol
\copyrightnotice

\begin{abstract}
%Modern manufacturing lines require high flexibility and reconfigurability to meet constantly changing production demands. Model-based Engineering provides powerful tools for engineers to design and rapidly redesign production lines to meet these demands. However, before final reconfiguration in the factory, additional tasks like simulations must be completed. Digital Twins enable engineers to efficiently monitor, simulate, and reconfigure production environments. This paper introduces a novel platform that automates the generation and deployment of manufacturing Digital Twins based on factory plan modeling using AutomationML. The DT closes the loop by providing a GAI-powered simulation scenario generator and automatic reconfiguration of the physical production line.
Modern manufacturing demands high flexibility and reconfigurability to adapt to dynamic production needs. Model-based Engineering (MBE) supports rapid production line design, but final reconfiguration requires simulations and validation. Digital Twins (DTs) streamline this process by enabling real-time monitoring, simulation, and reconfiguration. This paper presents a novel platform that automates DT generation and deployment using AutomationML-based factory plans. The platform closes the loop with a GAI-powered simulation scenario generator and automatic physical line reconfiguration, enhancing efficiency and adaptability in manufacturing.
\end{abstract}


\section{Introduction}
%Reconfigurability and flexibility are the cornerstone of the modern manufacturing environments. From the initial concept of Reconfigurable Manufacturing Systems (RMS) to Industry 4.0 and 5.0, the digitization of industrial systems aims to provide production lines with the ability to quickly and safely adapt to the dynamic production requirements~\cite{yelles2021reconfigurable}. The engineers responsible for the change to the factory shop floor are facing two major challenges: a) to accurately design the new production line and b) to deploy the change to the real manufacturing environment. The first challenge deals with the formal design of the products and the processes, utilizing tools such as modeling and simulation. The second refers to the reconfiguration of the controllers that manage the machines and the other industrial devices. 
Reconfigurability and flexibility are essential in modern manufacturing. From Reconfigurable Manufacturing Systems (RMS) to Industry 4.0/5.0, digital transformation enables production lines to adapt swiftly to changing demands ~\cite{yelles2021reconfigurable}. Engineers face two key challenges: (a) designing new production lines accurately using modeling and simulation, and (b) deploying these changes to physical systems by reconfiguring machine controllers and industrial devices.

%Digital Twins (DT) come to provide a powerful tool in the hands of engineers to easily and effectively monitor, simulate and reconfigure the production environment~\cite{mylonas2021digital}. 
Digital Twins (DTs) empower engineers to efficiently monitor, simulate, and reconfigure production environments ~\cite{mylonas2021digital}. 
%They provide a range of functionalities such as real time monitoring, realistic presentation of the environment in a 3D virtual world and interaction with it, updated prediction models based on real data, advanced simulation and automatic reconfiguration of the cyber physical systems participating in the production~\cite{leng2021digital}. 
They enable real-time monitoring, 3D virtual environment visualization, predictive modeling with live data, advanced simulation, and automated reconfiguration of cyber-physical production systems ~\cite{leng2021digital}. %Nevertheless, the implementation and deployment of a DT in production environment is usually a high-effort and costly task.
However, implementing and deploying a DT in a production environment remains a complex, resource-intensive process.

%Modeling the manufacturing environment based on well-known standards is a common practice to the engineers who design, monitor and maintain the production line. These models, either composed in SySML or AutomationML or in another meta-model, formally describe the manufacturing environment infrastructure including machines, controllers, processes and products~\cite{rath2023towards}. 
Engineers commonly model manufacturing environments using established standards like SysML or AutomationML. These formal meta-models describe production infrastructure—including machines, controllers, processes, and products—to support design, monitoring, and maintenance ~\cite{rath2023towards}. 
%The proposed approach takes advantage of this formalization to extract the basic concepts and automatize the DT deployment process. Furthermore, the proposed platform, taking advantage of the ability of Generative AI techniques to create documents based on specific concepts, introduces a new approach of generating simulation scenario for new production processes that can be simulated in the DT. Finally, the proposed DT platform provides a solution on quick reconfiguration of the production process, closing the loop of the interaction between real environment and DT. 
The proposed approach leverages this formalization to extract key concepts and automate DT deployment. By harnessing generative AI, the platform creates simulation scenarios for new production processes, enabling predictive testing in the virtual environment. Additionally, it facilitates rapid production line reconfiguration, closing the loop between the physical system and its digital counterpart. The proposed framework is based on authors' previous work~\cite{alexakos2020iot} where they employed AutomationML to enable rapid, semi-automatic deployment of an IoT data collection platform in industry. 

The rest of the paper is structured as follows. Section II presents related work and technologies utilized, section III details the proposed Model-based DT generation approach, while section IV presents a relevant use-case. Finally, section V provides discussion and conclusions.

\section{Background and Related Work}

\subsection{Model-based engineering in manufacturing}
Model-based engineering (MBE) in manufacturing encompasses the use of domain-specific languages and meta-models to design, simulate, integrate, and operate production systems throughout their entire lifecycle. While various modelling languages exist, the most widely adopted in modern manufacturing include SysML, AutomationML and the Asset Administration Shell (AAS) Information Model~\cite{khabbazi2024plug}. Beyond these core languages, specialized standards support specific functions: PLCopen XML for controller programming, STEP AP242 for product data exchange, OPC UA for semantic communication, Modelica for multi-physics simulation, and BPMN for process workflow modelling. These modelling frameworks form the foundation for designing, developing and operating Digital Twins in industrial environments. SysML typically defines the DT's system architecture and requirements, specifying functional needs, performance constraints, and system boundaries for accurate representation ~\cite{wilking2022sysml, ghanbarifard2024toward}. AutomationML, on the other hand, offers a more granular specification of the automation engineering layer, creating comprehensive plant models that capture mechanical layouts, electrical configurations, and control logic~\cite{zhao2021automationml, binder2021automated, lehner2021aml4dt}. The AAS serves as both a data structure and  runtime framework for DTs, maintaining standardized digital representations that persist throughout the asset lifecycle, incorporating both static information (technical specifications, documentation) and dynamic operational data (sensor readings, performance metrics, maintenance records)~\cite{redeker2021towards,abdel2022asset}.

The proposed DT framework utilises AutomationML for its comprehensive modelling capabilities of manufacturing automation systems. Moreover, it incorporates OMG's Business Process Model and Notation (BPMN) for production process definition, supported by an orchestration engine that executes processes in the physical environment.

\subsection{Generation of Digital Twins}

%There are few research solutions proposed the last five years to face the challenge of automatic or semi-automatic generation of DTs in industry, most of them focusing in specific manufacturing domain. 
Over the past five years, limited research has addressed the challenge of automated or semi-automated DT generation in industrial settings, with most solutions focusing on specific manufacturing domains. In ~\cite{sierla2022roadmap} the authors proposed a roadmap towards a methodology for the semi-automatic DT generation in brownfield environments, with particular emphasis on identifying key research challenges to enhance industrial applicability of their methodology.  Behrendt et al. ~\cite{behrendt2025real} leverage data mining and machine learning techniques to analyse existing manufacturing data and generate DTs. An alternative approach \cite{dalibor2022generating} employs UML to model DTs focusing on self-adaptive DT generation, specifically targeting  low-code development platforms. Kaiser et al. ~\cite{kaiser2022model} proposed  a model-based approach for generating simulation models in reconfigurable manufacturing systems utilizing AutomationML, yet their approach lacks automatic reconfiguration capabilities.

\section{Digital Twin automatic generation}

\subsection{Digital Twin generation concept}

The proposed approach automates the end-to-end DT lifecycle, covering concepts from the modelling of the real manufacturing environment, to the generation, deployment and operation of its DT.

\begin{itemize}
    \item \textbf{Step 1: Production Line Modelling}: AutomationML is used for the formal definition of the production line, exploiting a free-to-use AutomationML Editor for this task. The comprehensive model must incorporate all the information regarding the critical manufacturing elements comprising industrial machines and infrastructure, internal logistics network and control devices, including their topology at the shop floor, their functional representations using 3D object functional definition (in our case, we use URDF files for robotic devices and .FBX-based Unity Prefabs for other machines) and the controllers' calls (in our case, ModBus Addresses). The fully specified outcome AML file serves as the foundational input for the two next parallel steps in the DT generation pipeline.
    \item \textbf{Step2a: Concepts extraction} An AML parser processes the AML file to extract key manufacturing concepts including which machines are located at the shop floor, what are their functionalities and abilities, which are their controllers, and their interfaces for retrieving data and invoking commands and functions. These extracted elements are structured into a simplified JSON representation that serves as the configuration basis for the subsequent DT platform initialization.
    \item \textbf{Step2b: Virtual World Creation} 
    %The AML file and the control code (i.e. PLCs or Robotic Controllers) are used as an input to a script that creates the Virtual Twin of the production line. The base of this Virtual Work is a platform, implemented on top of Unity 3D Engine, which parses the AML and deploy in a virtual 3D space the industrial components including their functionalities. Further, virtual machines are instantiated for the emulation of the controllers of the industrial machines. OpenPLC and ROS empowered virtual machines are used for emulating the PLCs and the Robotic Controllers.
    The AML file and control logic (PLCs/Robotic Controllers) serve as inputs to an automated Virtual Twin generation process. Built on the Unity 3D Engine platform, the system parses the AML model to reconstruct the production environment in a virtual 3D space, complete with functional industrial components. For control emulation, the solution implements virtual machines running OpenPLC (for industrial controllers) and ROS (for robotic systems), accurately replicating the behaviour of physical automation devices.
    \item \textbf{Step 3: Digital Twin Platform Configuration} The extracted AML configuration data is imported in the DT platform, initializing all the parameters of the system for data collection, monitoring and visualization to the user. The platform automatically generates data structures in the ThingsBoard framework, assuming the role of data collection, being an IoT platform that supports multiple industrial protocols, such as ModBus, MQQT, and HTTP. At this stage, the Digital Twin establishes bidirectional connectivity with both physical assets and virtual representations, enabling real-time data streaming to the virtual word, and interactive visualization through graphical interaction interfaces.
    \item \textbf{Step 4: Industrial Process Generation} This step focuses on the generation of diverse production scenarios to be simulated in the virtual environment. This generation process is supported by a Large Language Model (LLM), which is fine-tuned for this purpose. The LLM leverages the production line's industrial capabilities to create alternative production processes scenarios, formally described in BPMN. These generated scenarios undergo virtual simulation and performance evaluation within the DT environment before deployment consideration.
    \item \textbf{Step 5: Closing the loop} At the final step, the validated production process, that has been successfully tested in the virtual environment, is deployed to the physical shop floor. The DT platform's integrated BPMN engine orchestrates the execution by dynamically coordinating the real control systems according to the prescribed process flow, completing the digital-physical cycle.   
\end{itemize}

This structured methodology enables rapid and reliable design and deployment of fully functional DT for production lines. The next subsections present the technical implementation aspects of the components presented in the aforementioned steps.   

\subsection{Model-based Digital Twin architecture}

\begin{figure}[ht]
    \includegraphics[width=\columnwidth]{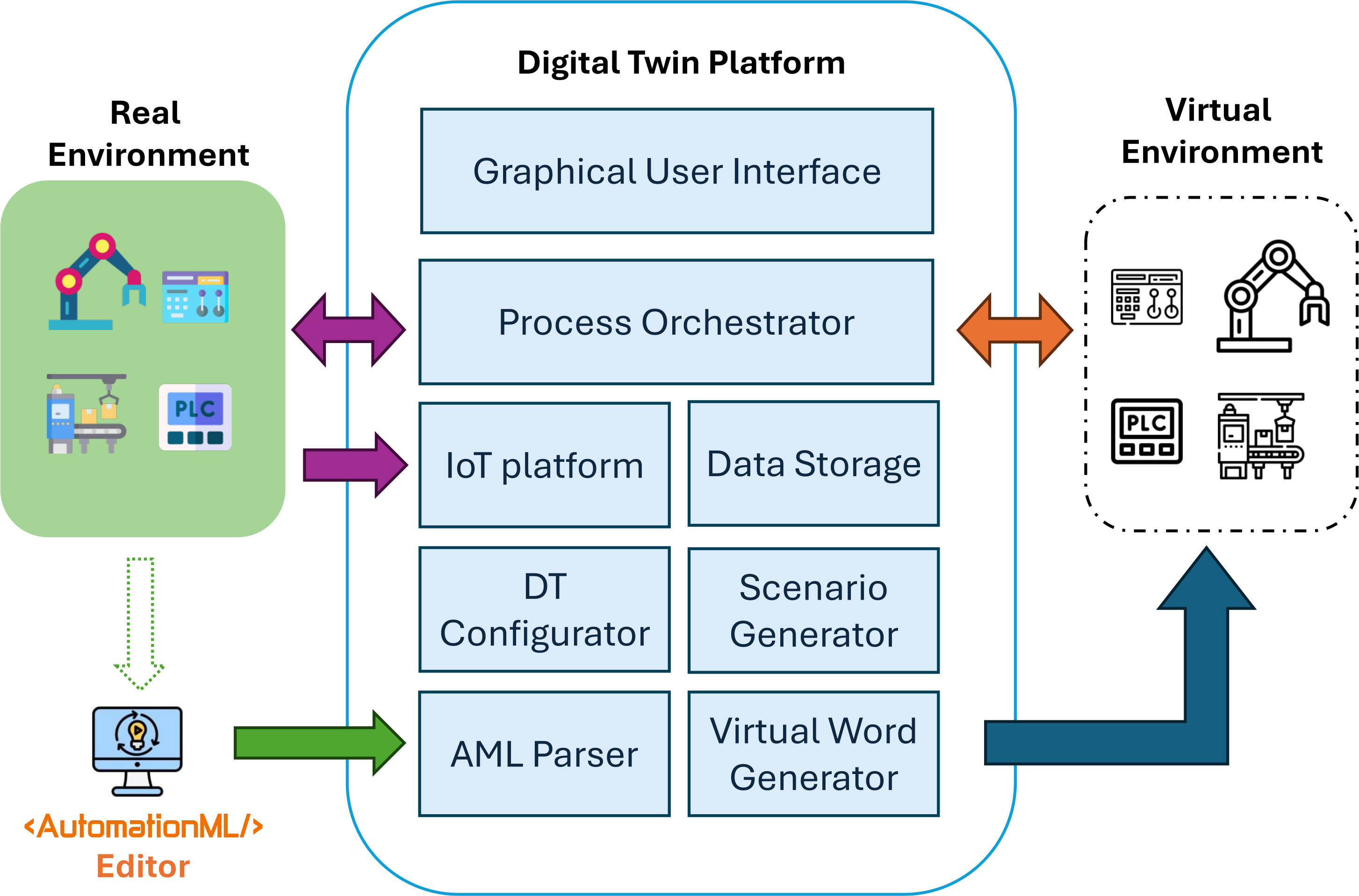}
    \caption{Digital Twin Architecture  (\textit{This image has been designed using resources from Flaticon.com})}
    \label{fig:dtarchitecture}
\end{figure}

A prototype of the DT platform has been implemented for supporting the generation conceptualization presented in the previous section. The basic modules of this platform are depicted in Fig. ~\ref{fig:dtarchitecture}.
The real manufacturing environment, containing machines, sensors, actuators and controllers, is initially modelled following the AutomationML notation utilizing the AutomationML tool. The produced AML file includes the description of the industrial components of the production line, using a hierarchical and interconnected model. This file serves as a comprehensive digital blueprint, encoding all required real-world concepts for DT generation, and constituting the foundational input for DT platform initialization and operation. The DT platform consists of the following functional units:
\begin{itemize}
    \item \textit{AML Parser} is a script which parses the AML file and exports  key manufacturing data, including shop floor equipment and systems, their topology, functional capabilities, controller command interfaces.
    \item \textit{Virtual Word Generator} creates a 3D digital replica of the physical environment in Unity,  instantiating controller virtual instances, emulating real-world counterparts.
    \item \textit{DT Configuration} configures \textit{Process Orchestrator} by initializing its operational parameters and deploying communication plug-ins for real-virtual controller interaction. Moreover, it establishes IoT platform data structures and messaging interfaces of communication modules to enable real-time data streaming from physical controllers.
    \item \textit{Scenario Generator} utilises Generative AI to synthesize potential production process scenarios based on the definition of the real production process. These AI-generated scenarios are formalized in BPMN with industrial devices serving as actors and control routines defining tasks. 
    \item \textit{IoT platform} leverages the ThingsBoard open source platform to establish real time connectivity with physical sensors, gateways and controllers enabling continuous data collection.  
    \item \textit{Data Storage} employs a hybrid database architecture, combining a NoSQL database (Cassandra DB) and a relational database (Postgres SQL) to store essential data, including collected telemetry from the physical world.
    \item \textit{Process Orchestrator} utilizes an embedded BPMN engine to coordinate production processes across both physical and virtual environments. During execution, it dynamically interacts with field devices by issuing control commands and polling sensor data streams.
    \item \textit{Graphical User Interface} is web-based providing comprehensive control over the DT, enabling users to monitor live data, generate and simulate process scenarios, and deploy validated changes directly to physical shop floor.
    %application allowing the users to interact and use the functionalities of the DT, from monitoring the production process, to generating scenarios, simulating in the virtual environment or even change the process in the real shop floor.       
\end{itemize}

The following subsections detail the platform's module functionalities, covering both deployment and operation.

\subsection{Manufacturing environment modelling}
The manufacturing environment is formally modelled in AutomationML (AML) through a structured representation of both physical assets and control systems. The AML schema organizes the factory into two interrelated hierarchies:

Physical Hierarchy:
Nested decomposition of assets, from factory level down to individual components, including:
\begin{itemize}
\item Production lines and workcells,
\item Machines (e.g., robots, CNC systems),
\item Sensors, actuators, and auxiliary equipment.
\end{itemize}
All entities are annotated with metadata defining functional properties, communication interfaces, and topological relationships.

Control Hierarchy:
A logical representation of automation resources, where PLCs, industrial PCs, and their software functions (e.g. motion control, state machines) are explicitly linked to the physical assets they govern. This mapping ensures traceability between cyberphysical components and their real-world counterparts.

\subsection{Digital Twin deployment}
System architecture is centred around a dynamic interplay between the Process Orchestrator, acting as the central coordination and intelligence hub, and a suite of protocol-specific middleware services, serving as the crucial abstraction layer. Middleware services are designed as modular, containerized applications, with each service variant engineered to communicate with a specific type of machine controller, such as those adhering to ROS or Modbus. Crucially, these services can seamlessly interact with both the physical hardware on the factory floor and the virtualized controllers within the Unity simulator, enabling a standardized command and event-driven communication model based on a RESTful HTTP API.

The configuration and deployment process is fully automated and based on the ingestion of the JSON file containing the complete inventory of available machines, their functional capabilities and communication protocols. With this information, corresponding middleware service instances are deployed. During deployment, each service is configured with precise endpoint details needed to connect to its designated controller, either a physical device or a virtual counterpart within the simulated environment. Upon successful instantiation, each middleware service establishes a communication link with the Orchestrator, confirming its operational status and readiness to receive commands and report events.

Once this configuration phase completes, the system enters its operational state, capable of managing the factory in two distinct modes: a fully virtualized mode for executing experimental scenarios solely within the Unity simulator, or a tightly coupled mode where the digital twin mirrors the physical factory by reading live sensor data. The Process Orchestrator executes high-level logic derived BPMN scenarios. The Orchestrator issues commands to the appropriate middleware services, which services then translate these abstract directives into the native language of the target controller. Throughout this process, the middleware provides a bidirectional flow of information, reporting back significant lifecycle events (such as task completion or errors) to the Orchestrator. In parallel, these services continuously stream low-level sensor telemetry via MQTT to a dedicated ThingsBoard installation. IoT Platform is leveraged for the live visualization of operational parameters on configurable dashboards, logging of historical data for performance analysis and diagnostics, and implementation of rule-based alerts for proactive fault detection.

\subsection{Process Generation \& Closing the loop with the physical environment}
%Describe how we generate the new processes, the ffedback loop from the user. (two-three paragraphs). 
%In the next phase four functional components of the proposed system, Scenario Generator, AML Parser, Process Orchestrator, and Graphical User Interface are combined to dynamically generate manufacturing processes, simulate them in the DT and with validation from a process engineer, deploy them in the real manufacturing environment, the Physical Twin.
The system's four core components—Scenario Generator, AML Parser, Process Orchestrator, and GUI—collaborate to automatically create manufacturable processes, digitally validate them via the DT, and deploy engineer-approved workflows to the Physical Twin.

\begin{figure}[ht]
    \includegraphics[width=\columnwidth]{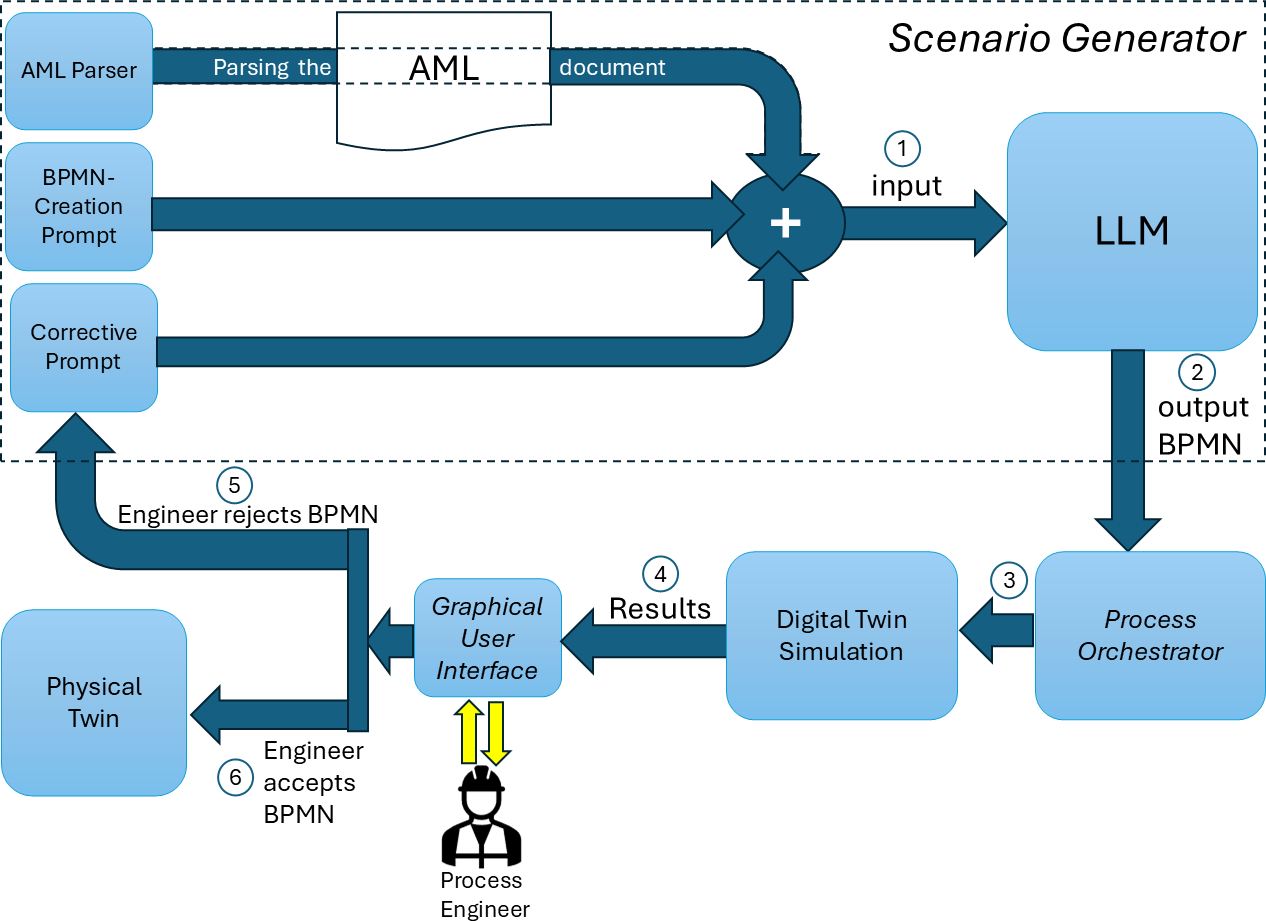}
    \caption{Business Process Generation depicting the transitions between various stages of the methodology}
    \label{fig:ProcGen}
\end{figure}

Figure \ref{fig:ProcGen} shows the scope of the process generation loop. In this instance, the Scenario Generator encompasses three different factors, which shape the input prompt to the LLM, the main component of the Scenario Generator. The information stemmed from the AML parser is combined with the BPMN-Creation Prompt, a prompt that explicitly informs the LLM to adhere to the BPMN standard when it generates it in XML format. The final component of the input prompt, is the Corrective Prompt which is provided by a human supervisor, denoting specific requirements or providing corrections or alteration suggestions to the LLM regarding the BPMN generated in the previous iteration (Transition-TR 1 in  Fig.\ref{fig:ProcGen}). In TR 2 the LLM produces a plain BPMN file, which is passed on to the Process Orchestrator for validation and pre-prossesing. Simulation ensues (TR 3), which accurately describes the process from start to finish, capturing the movement of machinery and other actors, such as robots and human personnel, as well as constantly updating relevant signals to PLC and other involved controllers (TR 4). A process engineer supervises the simulation as it's unravelling. The engineer, with the simulation results as feedback, can decide at any point that the process produces the desired result, and can opt to modify the Corrective Prompt and regenerate an updated BPMN (TR 5). In case simulation results are validated by the engineer, process is stored to Process Orchestrator, being available to execute in the physical environment.

\section{Use case}
The validation of the proposed methodology and system has been conducted on a production line demonstrator. This demonstrator is emulating a physical production environment utilizing industrial machine miniatures. The validation process, presented in the following sections, includes all steps of the model-based DT generation, concluding to the successful DT operation providing real-time monitoring, simulations and reconfiguration of the production line.

\subsection{Production process demonstrator}
\begin{figure}[ht]
    \includegraphics[width=\columnwidth]{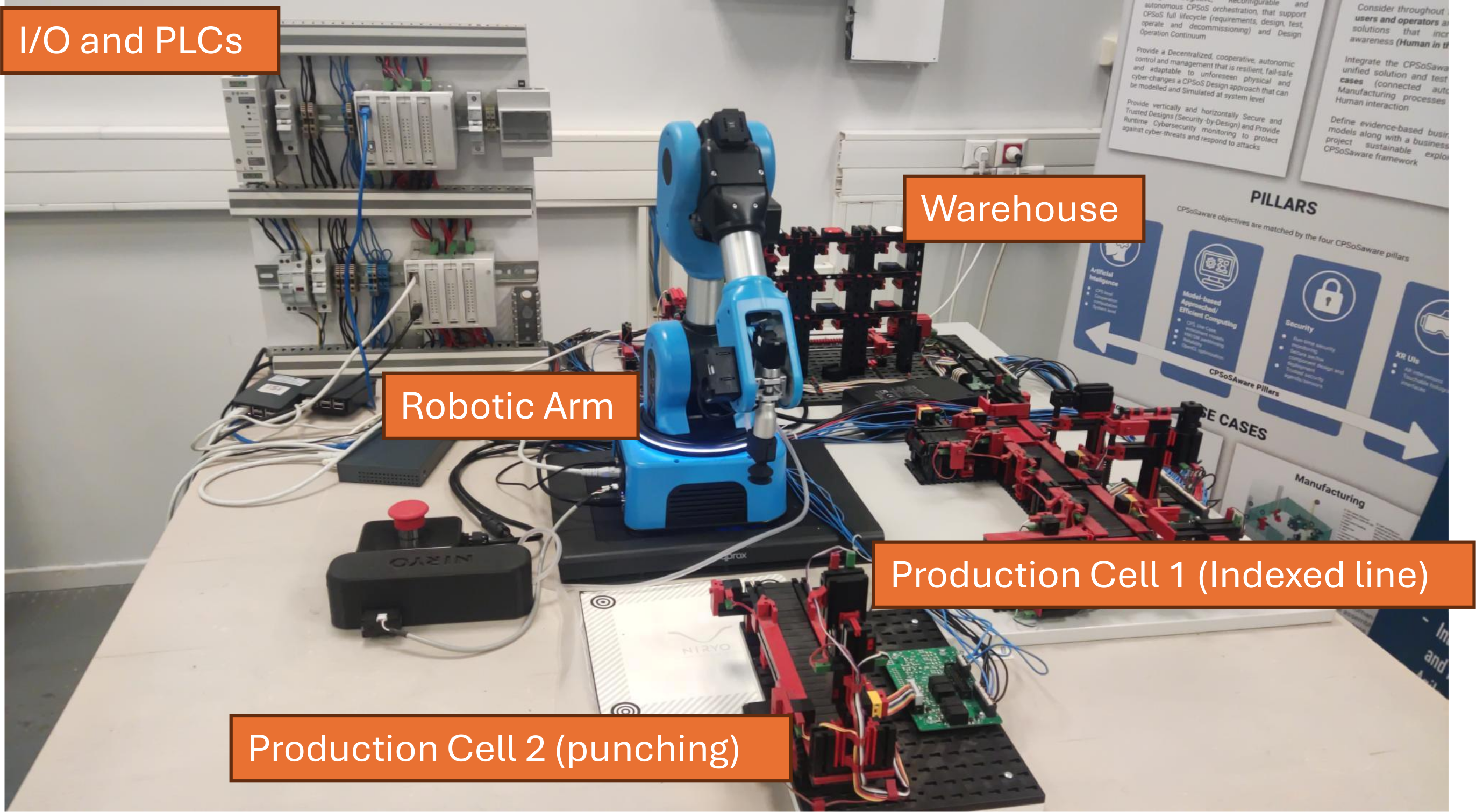}
    \caption{Physical Production line Demonstrator Environment}
    \label{fig:PhysicalEnvironment}
\end{figure}

%description of the real environment, we translate from greece 4.0 deliverables (two-three paragraphs)
The production line demonstrator (Fig. ~\ref{fig:PhysicalEnvironment}) consists of five industrial components: a) A Niryo Ned2 robotic arm executing internal logistics tasks, b) an Automated High-Bay Warehouse for storing the intermediate products, c) a Punching Machine with Conveyor Belt representing one of the production shells, d) an Indexed Line with two Machining Stations representing a second production shell (components b,c,d provided by Fishertech), and e) I/O managing units, robot controller and control PLCs.
 
Machine electrical signals are transmitted through cables to an Arduino-based I/O slave unit (provided by Industrial Shields), which transforms these signals into Modbus TCP packets for transmission to an OpenPLC installation on a Rasberry Pi. OpenPLC serves as an open-source industrial controller that complies with IEC 61131-3 standards and includes an integrated development environment that enables users to create programs using Program Organization Units (POUs). The controller operates using the Modbus protocol over a TCP/IP network stack. Furthermore, robot control is done through a gateway for ROS control operating system installed on a Rasberry Pi.

\subsection{Physical world environment modelling}

The physical-world manufacturing environment is comprehensively captured in an AutomationML (AML) model through a dual-structured hierarchy that integrates both physical and cyber components. 

At the core of the AML model lies a detailed decomposition of physical assets, encompassing key operational zones such as the Factory, Warehouse, and Processing Area. These are further broken down into constituent machines and components, including: a)\textit{ Material handling systems}, such as conveyors with photoelectric sensors and motors, b) \textit{Processing units}, like the milling machine, drilling station, and punching machine, c) \textit{Robotic manipulators}, such as articulated robotic arms, and \textit{Auxiliary equipment}, like counters, switches, and linear actuators. Each physical entity is described via \texttt{SystemUnitClasses} and \texttt{RoleClasses}, and is associated with external interfaces (e.g., \texttt{IPhotoSensor}, \texttt{IDCMotor}, \texttt{IPLCSignalInterface}) to define connectivity and functional roles. These interfaces also facilitate interaction with the automation layer, enabling consistent integration of hardware with control logic.

In addition to the physical hierarchy, the AML model defines a control hierarchy that captures logic, data flow, and supervisory behaviour within the manufacturing system. This includes: a) \textit{Programmable Logic Controllers (PLCs)} such as \texttt{PLC1}, which interface with physical devices to manage actuation and sensing operations, b) \textit{ROS-based controllers} (e.g., \texttt{ROS\_RasPI}) coordinating robotic command execution and sensor feedback, c) \textit{Programmed control functions}, encapsulated as internal AML elements (e.g.\texttt{LoadFromWarehouse}, \texttt{StoreToWarehouse}, \texttt{MillAndDrill}, \texttt{Stamp}, \texttt{RobotCommand}), which describe high-level operational tasks. Each function is explicitly bound to its corresponding hardware unit through reference links and interface declarations (e.g., \texttt{IPLCMachineInterface}, \texttt{IOPCInterface}, \texttt{IUnityInterface}). This facilitates a \textit{traceable cyber-physical mapping}, ensuring that every control function is contextualized by its target physical asset.

Furthermore, the AML structure reflects a modular and reusable modelling approach, where machines such as conveyors, motors, and photo sensors are instantiated from generalized component libraries (e.g., \texttt{SimpleComponents/Conveyor}, \texttt{SimpleComponents/PhotoSensor}). This ensures consistency across system components as well as extensibility.

\subsection{Deploying DT}
\begin{figure}[ht]
    \includegraphics[width=\columnwidth]{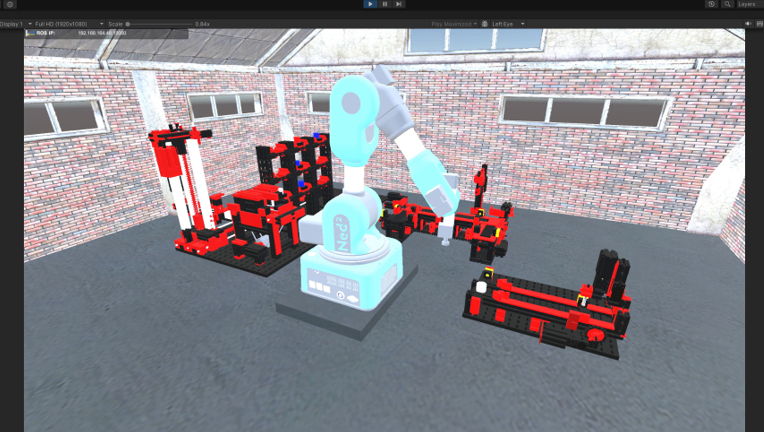}
    \caption{Digital Twin of the production line}
    \label{fig:dtvirtual}
\end{figure}

The DT deployment process starts with the parsing of the created AutomationML file. The Orchestrator ingests the model and identifies the four primary assets: an indexed machine with a milling and a drilling station, an automated warehouse, a punching machine, and a robotic arm and related information. Subsequently, it provides this information to the Unity 3D engine, which instantiates the virtual twin by rendering the assets' 3D prefabs and starting the virtual machines of the virtual controllers (Fig. ~\ref{fig:dtvirtual}). 
Furthermore, by analysing the specifications, it determines that these virtual assets require distinct communication protocols, flagging the robotic arm for a ROS interface and the other three machines for control via Modbus TCP.  Based on this analysis, the necessary middleware services to control the simulated factory are configured. A dedicated ROS service is launched for the robotic arm, and a Modbus TCP service is created for the PLC-controlled machines. For this virtual deployment, the Orchestrator configures the services with the network endpoints of the controllers running within the simulation environment. Once launched, the services establish communication with their respective virtual counterparts and report back to the Orchestrator. This confirms that the entire digital twin of the cell is online, fully connected and ready for virtual commissioning or the execution of experimental manufacturing scenarios.

\subsection{Digital Twin for process simulation and reconfiguration}

As described before, the Process Generator loop is employed to formulate the process, that will be tested in the DT, and then if deemed sufficient, will be applied to the real environment. To initiate the loop, all components must be functional and in order.  For the LLM one of the two most prominent frontier models for code generation are tested, ChatGPT-4o and claude-3-7-sonnet. For both models the generation time, with the given input prompt, is comparable: about thirty seconds. Their results are also close, but Claude tends to require fewer corrective iterations to achieve the desired BPMN process. 
%A combination of these two was used, where ChatGPT is able to provide coherent BPMN files, without too many corrections, and Claude will refine adding more complex and robust code such as BPMN Lanes and Pools.
A combination of these two was used, with ChatGPT providing coherent BPMN files without too many corrections, while Claude offering refinement adding more complex and robust code, such as BPMN Lanes and Pools.

\begin{figure}[ht]
    \includegraphics[width=\columnwidth]{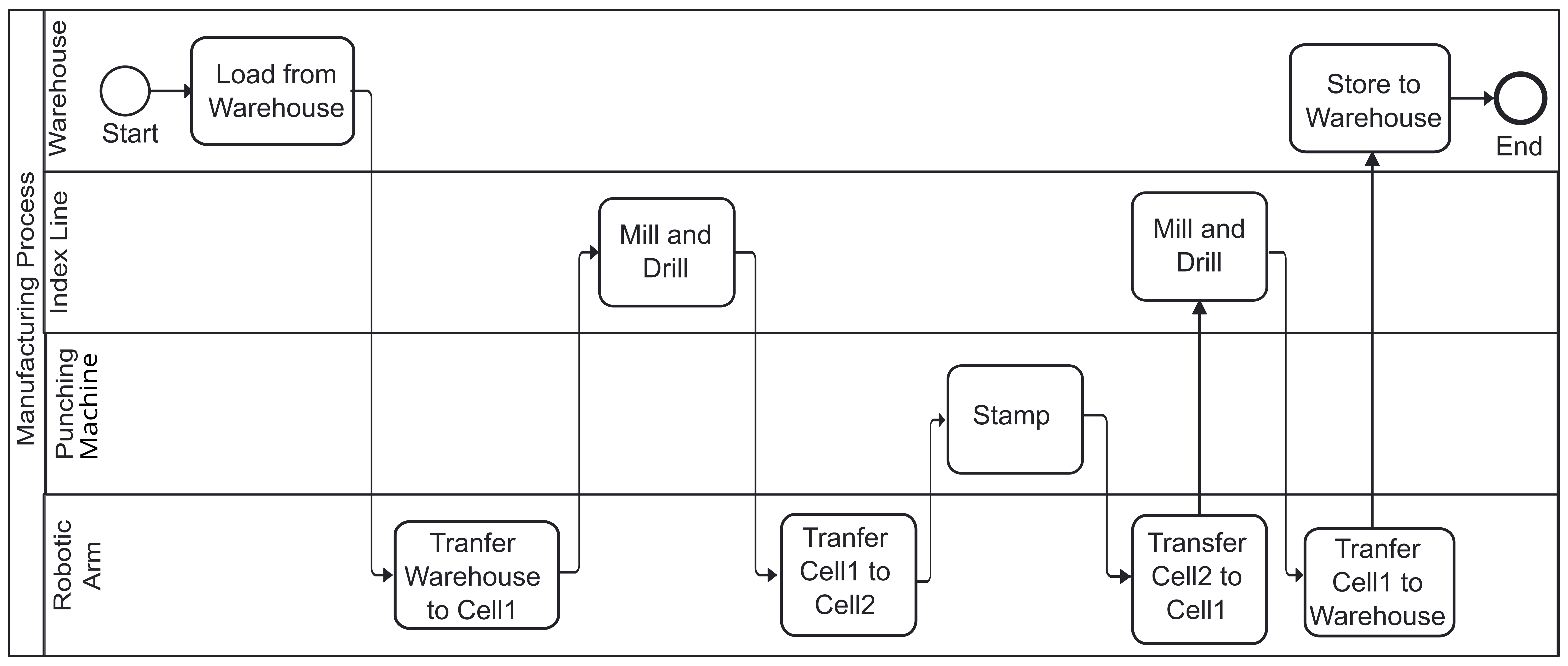}
    \caption{Generated Process}
    \label{fig:BigProcess}
\end{figure}

Two cases were tested with the Scenario Generator. The first process prompt is described as "take the item from the warehouse, place it on the punching machine, stamp, place on index line, mill \& drill, and finally store it back in the warehouse". The second process is described as "Take the item from the warehouse, place it on the index line, mill \& drill it,  place on the punching machine, stamp it , place it again on the index line, mill \& drill again, then finally return to the warehouse". The result of the second is presented in Fig. \ref{fig:BigProcess}. A finding during the experiments was that the simpler the process is, the less corrections and validation iterations with the engineer are needed. The final step is the execution of the BPMN from the Process orchestrator. The integrated BPMN engine is able to invoke the necessary commands in the physical assets' controller, following the same behaviour of invoking commands to the DT, ensuring the correct execution of the process to the real production line.

\section{Conclusions}
%Model-based Engineer provides the appropriate tools for the diligent and formal definition of the industrial systems, products and processes comprising a manufacturing environment. The proposed approach leverages this formalization and provides a platform for generating and deploy DTs based on AutomationML rationalization of a production line. Further, it is empowered with a Generative AI engine for generating simulation scenarios of production processes which are validated in the Virtual Twin, before automatically deployed in the real environment. The validation through use-case was successful and also highlighted the weaknesses of the proposed approach, which will be the focus of the future work. Complex manufacturing lines feature dozens of machines, where the utilization of simple LLM techniques became more challenging as they have to deal with more complex concepts and more large text segments. 
Model-based Engineering provides the methodological foundation for rigorous formalization of industrial systems, encompassing manufacturing assets, production processes, and product definitions. The presented platform leverages this structured approach through AutomationML-based digital representations to enable systematic DT generation and deployment. A key innovation lies in the integration of Generative AI capabilities, which automates the creation of feasible production scenarios that undergo virtual validation before being implemented in the physical environment. While successful implementation through industrial use-cases has demonstrated the approach's viability, the evaluation also revealed critical limitations, particularly when scaling to complex manufacturing systems. As production lines grow to incorporate dozens of interconnected machines with intricate dependencies, current LLM techniques face substantial challenges in processing the resulting complexity of concepts and the exponential growth of contextual information. These identified constraints, stemming primarily from the semantic understanding limitations of conventional language models in industrial contexts, will form the primary focus of forthcoming research efforts aimed at enhancing AI robustness for manufacturing applications.

\bibliographystyle{IEEEtran}
\bibliography{seeda_greece4.0}

\end{document}